\documentclass[12pt]{iopart}

\usepackage{iopams}  

\usepackage{graphicx,epsfig}
\usepackage{color}
\usepackage[normalem]{ulem}

\newcommand\be{\begin{equation}}
\newcommand\ee{\end{equation}}
\newcommand\bea{\begin{eqnarray}}
\newcommand\eea{\end{eqnarray}}

\newcommand\qq{{\bf q}}
\newcommand\rr{{\bf r}}

\newcommand\ex{{\bf e}_x}
\newcommand\ez{{\bf e}_z}
\newcommand\er{{\bf e}_r}
\newcommand\epsi{{\bf e}_\psi}

 \newcommand{\sinc}{{\rm sinc}}

\begin{document}

\title[Differential dynamic microscopy of helical and oscillatory microswimmers]{Helical and oscillatory microswimmer motility statistics from differential dynamic microscopy}

\author{Ottavio A. Croze $^1$, Vincent A. Martinez $^2$,  Theresa Jakuszeit $^{1}$, \\Dario Dell'Arciprete $^{3}$,
	Wilson C. K. Poon $^2$\\ and Martin A. Bees $^4$}
	
\address{$^1$ Cavendish Laboratory, University of Cambridge, J.J. Thomson Avenue, Cambridge, CB3 0HE, United Kingdom}
\address{$^2$ School of Physics and Astronomy, University of Edinburgh, Peter Guthrie Tait Road, Edinburgh EH9 3FD, United Kingdom}
\address{$^3$ Dipartimento di Fisica, Universit\`{a} di Roma ``Sapienza'', Rome, I-00185, Italy}
\address{$^4$ Department of Mathematics, University of York, York YO10 5DD, United Kingdom}
\ead{oac24@cam.ac.uk, vincent.martinez@ed.ac.uk}
\vspace{10pt}
\begin{indented}
\item[Manuscript draft date: \today]
\end{indented}

\begin{abstract}
The experimental characterisation of the swimming statistics of populations of microorganisms or artificially propelled particles is essential for understanding the physics of active systems and their exploitation.  Here, we construct a theoretical framework to extract information on the three-dimensional motion of micro-swimmers from the Intermediate Scattering Function (ISF) obtained from Differential Dynamic Microscopy (DDM). We derive theoretical expressions for the ISF of helical and oscillatory breaststroke swimmers, and test the theoretical framework by applying it to video sequences generated from simulated swimmers with precisely-controlled dynamics. We then discuss how our theory can be applied to the experimental study of helical swimmers, such as active Janus colloids or suspensions of motile microalgae. In particular, we show how fitting DDM data to a simple, non-helical ISF model can be used to derive three-dimensional helical motility parameters, which can therefore be obtained without specialised 3D microscopy equipment. Finally, we discus how our results aid the study of active matter and describe applications of biological and ecological importance.
\end{abstract}

%\maketitle

\section{Introduction}

The behaviour of swimming microorganisms and artificially propelled microscopic particles, collectively termed ``microswimmers,'' is of both fundamental and practical interest.  On the one hand, suspensions of microswimmers reveal qualitatively distinct statistical mechanics \cite{Cates2012, Elgeti2015, Bechinger2016} and fluid mechanics \cite{PedleyKessler92, GuastoRusconiStocker12, LaugaPowers09, Elgeti2015, Bechinger2016} from those of passive colloids. On the other hand, a quantitative understanding of microswimmer dynamics opens up exciting new possibilities for active material engineering \cite{Bechinger2016, Arlt2018} and microbial biotechnologies \cite{BeesCroze14}, as well as for microbe-dependent environmental and climate science, such as in marine plankton population dynamics \cite{DurhamStocker12, GuastoRusconiStocker12}. 

Microorganisms and artificial swimmers typically are observed to swim along trajectories of a helical nature \cite{Shengetal07, Campbell2017, Su16018}. The helical motion arises from body rotation about an axis that differs from the swimming direction \cite{Bearon13, Campbell2017} and is the expected outcome of systems that lack perfect symmetry.  For biological swimmers such as microalgae \cite{Shengetal07, Drescher2009} or spermatozoa \cite{Su16018} the rotation originates from torques caused by non-planar flagellar motion \cite{Bearon13, Hopeetal16}.  In the case of microalgae, the rotation is observed to have a biological role in that it allows cells to sample the light environment and move towards regions that are photosynthetically optimal (phototaxis) \cite{FosterSmyth80, Arrieta2017}. For artificial microswimmers such as catalytic Janus particles the rotation is likely due to a combination of body and coating imperfections \cite{Campbell2017}.

Measurements of the physical characteristics of swimmers permit the parameterization and improvement of theoretical models of active matter, which, if successfully predictive, can be used for innovative (bio)engineering design \cite{BeesCroze14, Bechinger2016}. Significantly, even in the absence of a theoretical framework, measurements of motility statistics also allow one to make direct inferences on the biological, ecological and biotechnological behaviour of microswimmers. For example, motility patterns, including helical swimming, and associated motility statistics change when heterotrophic microalgae prey on smaller phytoplankton \cite{Shengetal07}.
Therefore, it is crucial to be able to characterise three-dimensional swimming motions for statistically significant population sizes.

Single-particle tracking \cite{CrockerGrier96} and ensemble-averaged techniques, such as dynamic light scattering (DLS) \cite{BernePecora}, have been until recently the main techniques used to probe the spatio-temporal dynamics of particle suspensions. Particle tracking in video microscopy, developed for the characterisation of passive colloidal dynamics \cite{CrockerGrier96}, has been widely applied to study active systems, biological \cite{RafaiJibutiPeyla10, Hopeetal16} and synthetic \cite{Campbell2017, Kurzthaler2018}. However, the characterisation of three-dimensional motion, such as helical swimming, with standard imaging microscopy is limited by the tracking depth of the microscope \cite{CrockerGrier96}. Specialised microscopy apparatus and image processing algorithms are required to extract three-dimensional motility information, such as multiple cameras \cite{Crenshaw2000, Boakesetal00, Drescher2009, Orchard2016}, exploitation of optical phase information in phase-contrast microscopy \cite{Taute2015}, digital holographic microscopy \cite{Shengetal07, Farthing2017} or `Lagrangian microscopes' \cite{BergBrown, Darnige2017, Liu2014}. However, these techniques can be limited either in statistical accuracy and/or to low particle concentrations. For example, the use of multiple cameras is limited to relatively dilute samples because cross-correlation of camera outputs becomes challenging at high concentrations (due to particle trajectory overlaps) \cite{Drescher2009}. %Statistics are thus limited in individual experiments. 

In DLS, fluctuations in the light scattered from a sample are collected in the far field at a given scattering vector $\mathbf{q}$, and analysed to infer the microscopic dynamics. DLS delivers statistical information for the dynamics of colloidal samples in three dimensions and plays a crucial role in the study of passive soft matter \cite{BernePecora}. Importantly, DLS allows one to measure the Intermediate Scattering Function (ISF), also known as the Dynamic Structure Factor. The ISF is the Fourier component of the probability density function of particle displacements at a given time. It thus encodes full statistical information about the particle dynamics at a given length scale $l=2\pi/q$, with $q=| \mathbf{q}|$, and delay time $\tau$. While the potential for the application of DLS to suspensions of motile microorganisms was recognised early (e.g.~for bacteria \cite{Boon1974}, microalgae \cite{Racey1981, Racey1983} and sperm \cite{Craig1982}), standard DLS is restricted to large scattering angles, corresponding to small length-scales or large $q=|\mathbf{q}|$ values. At these small length-scales, many processes such as ballistic, rotational, and oscillatory motions all contribute to the ISF.  Thus extracting dynamical information becomes impractical. Therefore, DLS is not suitable for the study of microswimmers. In particular, because of its limitation to motions on small scales, DLS studies \cite{Schaefer1974, Holz1978, Craig1982} did not succeed in obtaining information about helical swimming trajectories, as this requires probing larger scale dynamics.

The discovery and development of differential dynamic microscopy (DDM; see \cite{Cerbino2017} for a review on recent developments) has made the dynamics of active systems amenable to being probed by standard imaging microscopy \cite{Wilsonetal11, Martinezetal12}. DDM yields the ISF and is particularly suited to low optical resolution imaging microscopy with a large field of view, thus giving access to the particle dynamics over large length-scales, i.e. more than one order of magnitude larger than DLS. Additionally, DDM is not restricted to low particle concentrations \cite{SchwarzLinek2016} and thus can more easily provide statistically significant information for dense microswimmer suspensions.

DDM has been applied to a range of microorganims to extract key motility parameters (including bacteria \cite{Wilsonetal11, Martinezetal12, Cerbino2018}, algae \cite{Martinezetal12} and spermatozoa \cite{Martinez388777}). For bacteria, DDM has also been used to clarify the interaction between motile and non-motile cells \cite{Jepson2013}, characteristics of swimming in a polymer solution \cite{Martinez2014}, and dynamics of concentrated suspensions \cite{Lu2012}. Also, it has been employed to study biological active matter (e.g.~\cite{SchwarzLinek2016, Rosko2016}) and artificial swimmers in quasi-two-dimensional geometries \cite{Kurzthaler2018}. However, despite the fact that many of the above artificial and biological swimmers swim helically, to the best of our knowledge, no theoretical expression for the ISFs of helical swimmers has been derived to allow the use of DDM to study their full motion. 

In this work we derive the ISF for swimmers with helical trajectories combined with progressive back-and-forth body motion. The latter is included so that the ISF can be used to describe biflagellate algae, such as the model species {\it Chlamydomonas reinhardtii}, which propel themselves by beating flagella with a breaststroke motion \cite{Goldstein2015}. We then derive approximations to the ISF that facilitate extraction of helical and breaststroke swimming statistics from DDM data. The accuracy of these approximations is assessed with video sequences generated from simulated microswimmers. Finally, we discuss how our analysis suggests a new method to extract helical swimming parameters using DDM with standard microscopy setups and simple ISF models. Our method should allow the experimental study of more concentrated suspensions of active swimmers than afforded by current 3D methods. 

\section{Theory: intermediate scattering function and approximations\label{sec:theory}}

The ISF for independent (non-interacting) swimmers is given by: 
\begin{equation}\label{eq:ISF}
f(q,\tau) = \left\langle e^{i {\bf q}\cdot  \Delta {\bf r}_j(t+\tau)}\right\rangle
\end{equation}
where $\Delta \rr_j(t+\tau)=\rr_j(t+\tau)-\rr_j(t)$ is the displacement of swimmer $j$, $\tau$ the delay time, {\bf q} is the wavevector, with magnitude $q=|\mathbf{q}|$, probing the dynamics at a length-scale $l=2\pi/q$, and angled brackets denote averages over time $t$, and all swimmers and wavevector direction $\mathbf{q}/q$. The position vector of a helical swimmer with an oscillating back-and-forth component is given by 
\be\label{eq:qDr}
\rr(\tau)=\rr_c+\delta\rr_b,
\ee
\begin{figure}[h!]
   \begin{center}
      \includegraphics[width=0.7\linewidth]{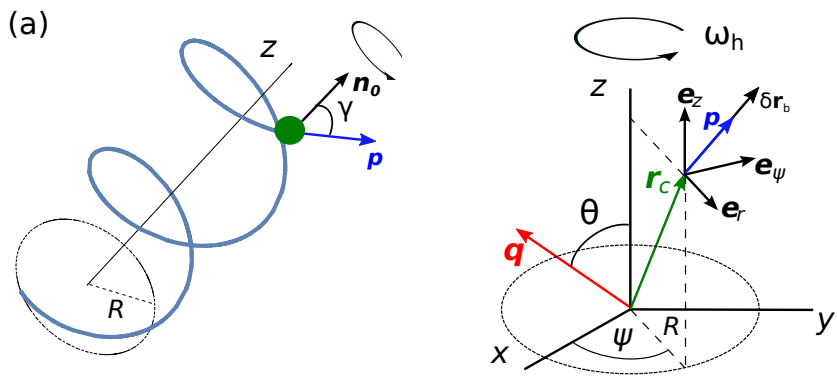}
      \caption{(a) Helical trajectory, of radius $R$, traced by a swimmer swimming in direction $\mathbf{p}$ and, due to internal torques (e.g. azimuthal components to the flagellar beat of biflagellate algae), rotating around the direction $\mathbf{n}_0$ making an angle $\gamma$ with $\mathbf{p}$. This direction coincides with the orientation of the traced helix. (b) The helix frame used to evaluate the Intermediate Scattering Function. The $z$-coordinate is aligned with the helix axis, around which the swimmer, with centre-of mass position $\mathbf{r}_c$, rotates with angular speed $\omega_h$, sweeping an azimuthal angle $\psi=\omega_h\tau+\phi_h$. Superposed on this motion along the instantaneous swimming direction $\mathbf{p}$ is a back-and-forth oscillation $\delta\rr_b$. The vector $\mathbf{q}$ is the scattering vector, as in dynamic light scattering.}.
      \label{fig:axes}
   \end{center}
\end{figure}
%F
where $\rr_c$ is the position of the centre around which the particle centre oscillates back-and-forth, with a diplacement $\delta\rr_b(\tau)$. Here and henceforth, we omit the swimmer index $j$ (e.g. $\rr_{c,j}\to\rr_c$) for clarity and with a view to later replacing sums over swimmers with integrals. The tip of the vector $\rr_c$ traces a helical path, and $\delta\rr_b$ models a back-and-forth oscillatory motion, such as the body rocking that results from the back-and-forth swimming of biflagellate algae like {\it Chlamydomonas spp.} \cite{Goldstein2015}, see figure \ref{fig:axes}. For each swimmer, we consider a reference frame with the $z$ axis coinciding with that of the helix, so that, adopting the cylindrical polar coordinate system $(r, \psi, z)$ shown in figure \ref{fig:axes}(b), we can decompose the helical motion as a superposition of translation along the helix axis and rotation around it
\be\label{eq:rH}
\rr_c=v_p\tau \ez + R\er,
\ee
where $v_p$ is the progressive speed, the projection of the cell velocity on the helical axis, and $R$ is the helical radius. The unit vector $\er$ rotates around the helical axis with angular speed $\omega_h=2\pi f_h$ (with $f_h$ the helical frequency), described by the azimuthal coordinate $\psi=\omega_h \tau+\phi_h$, where $\phi_h$ is a random phase, uniformly distributed in the interval $[0, 2\pi]$, added to ensure the helical rotations of different cells are not synchronised. The back-and-forth motion is along the instantaneous swimming direction
\be\label{eq:helicalspeed}
\mathbf{p}={\bf v}_p/v_{h}=(v_p \ez+\omega_h R\epsi)/v_{h}, \mbox{~~~where~~~} v_{h}=\sqrt{v_p^2+(\omega_h R)^2}.
\ee
This is the speed along the helical trajectory. The back-and-forth displacement can then be written as
%$\delta\rr_b=\delta r_b[(v/v_{h}) \ez +(\omega_h R/v_{h}) \epsi]$ 
\be\label{eq:drB}
\delta\rr_b=\delta r_b\left[\frac{v_p}{v_{h}} \ez + \frac{\omega_h R}{v_{h}} \epsi\right],
\ee
where
\be\label{eq:drBmag}
\delta r_b(\tau)=A_b \sin(\omega_b \tau+\phi_b).
\ee
Here, $A_b$ and $\omega_b=2\pi f_b$ are the back-and-forth oscillatory amplitude and angular speed (with $f_b$ the back-and-forth oscillatory frequency), respectively, and $\phi_b$ a random phase added to avoid synchronisation, as for the helical motion. 

With these approximations, the phase contribution to the ISF due to swimmers can be shown to be given by (see \ref{sec:appendixA})
\bea\label{eq:ISFalgaederiv2}
&&\eta\equiv\qq\cdot\Delta\rr(\tau) = q v_p \tau\cos\theta\\
&& + q A_b \frac{v_p}{v_{h}}\left[\sin\left(\omega_b\tau+\phi_b\right)-\sin\phi_b\right]\cos\theta\nonumber \\
&& + q R\left[\cos\left(\omega_h\tau+\phi_h\right)-\cos\phi_h\right] \sin\theta\nonumber\\
&& - q A_b \frac{\omega_h R}{v_{h}} \left[\sin\left(\omega_h\tau+\phi_h\right)\sin\left(\omega_b\tau+\phi_b\right)-\sin\phi_b\sin\phi_h\right]\sin\theta, \nonumber
\eea
where $\theta$ is the angle between $\qq$ and the helical axis along $z$. % and $v(v_{h})=\sqrt{v_{h}^2-(\omega_h R)^2}$ from (\ref{eq:helicalspeed}). 
The first term in equation (\ref{eq:ISFalgaederiv2}) is the contribution of progressive swimming, the second term corresponds to breastroke swimming oscillations, the third to helical rotation, and the forth is a cross-term reflecting the coupling between breastroke and helical swimming.

To evaluate the ISF given by (\ref{eq:ISF}) we need to average over all swimmers, which means averaging over all distributions of swimming parameters. To make analytical progress, henceforth we assume isotropic swimming, and consider a distribution $P(v_p)$ only for swimming speeds: all other swimming parameters are approximated as having single values. The ISF for the algae is then a multiple integral over the volume of the swimming velocity space ${\bf \Gamma_s}$, as well as over the random phases $\phi_h$ and $\phi_b$
\be\label{eq:IFDalgaederiv3}
f(q,\tau) =\left\langle e^{i \eta}\right\rangle \equiv\frac{\int\int d\phi_h d\phi_b\int_{\bf \Gamma_s} P(v_p) e^{i \eta}  d^3{\bf v}_p}{\int\int d\phi_h d\phi_b\int_{\bf \Gamma_s}  P(v_p)  d^3{\bf v}_p},
\ee
where normalisation ensures $f=1$ for $q=0$ or $\tau=0$. Upon substituting (\ref{eq:ISFalgaederiv2}) into (\ref{eq:ISF}) and choosing spherical polar coordinates for the velocity space integration we obtain
\be\label{eq:IFDalgaederiv3_new}
f (q,\tau) =\frac{1}{8\pi^2}\int_{0}^{2\pi}d\phi_b \int_{0}^{2\pi}d\phi_h \int_0^{\pi}\sin \theta  d\theta \int_{0}^\infty   P_s(v_p) e^{i\eta} d v_p,
\ee
where the isotropic speed distribution is defined as $P_s=4\pi P(v_p) v_p^2$. 

\subsection{Speed distribution transformations}

The reader will have noticed that the averaging over speed in the ISF integral (\ref{eq:IFDalgaederiv3_new}) is evaluated using $P_s(v_p)$, the distribution of progressive speeds $v_p$ (projected onto the helical axis), and not $P_s(v_{h})$, the distribution of along-helix speeds. In fact, either distribution can be used, as a simple change of variables in the integral demonstrates: $\int_{\omega_h R}^\infty h(v_{h})  P_s(v_{h}) d v_{h} = \int_0^\infty h(v_{h}(v_p))  P_s(v_p) d v_p$, where $h$ is a general function. This same result implies we can derive the along-helix motility statistics if we know the distribution of progressive speeds. Indeed, from equation (\ref{eq:helicalspeed}) it follows that the mean along-helix speed average is given by 
\be\label{eq:avg_inst_speedmom}
\overline{v_h}=\int_0^\infty [v_p^2+(\omega_h R)^2]^{1/2}\,P_s(v_p) dv_p,
\ee
where here and henceforth overbars denoted averages over the speed distribution. The speed average in equation (\ref{eq:avg_inst_speedmom}) provides a value of the along-helix speed given values of $\omega_h$, $R$, and the distribution $P_s (v_p)$. Similarly, recalling $P_s$ is normalised, it is easily shown that the second moment of the along-helix speed is given by $\overline{v_h^2}=\overline{v_p^2}+(\omega_h R)^2$. Using this relation and $\sigma_p^2=\overline{v_p^2}-\overline{v_p}^2$, the variance of the progressive speed distribution, the variance of the along-helix speed distribution can then be written as
\be\label{eq:inst_speedvar}
\sigma_h^2=\overline{v_h^2}-\overline{v_h}^2=\sigma_p^2+\overline{v_p}^2-\overline{v_h}^2+(\omega_h R)^2.
\ee
We will use transformations (\ref{eq:avg_inst_speedmom}) and (\ref{eq:inst_speedvar}) in the analysis of our simulated results, and in particular to deduce three-dimensional swimming parameters from two-dimensional data. These relations apply generally for swimmers with swimming distributions with finite first and second moments. In what follows, as in \cite{Wilsonetal11,Martinezetal12}, we consider swimmers with a Schulz distribution \cite{PuseyVanMegen84} of progressive swimming speeds, such that
\be\label{eq:schulz}
P_s(v_p)=\frac{1}{Z!}\left(\frac{Z+1}{\overline{v}_p}\right)^{Z+1}e^{-\left(\frac{Z+1}{\overline{v}_p}\right)v_p}v_p^Z, 
\ee
where $\overline{v}_p$ is the progressive mean speed and $Z=(\overline{v}_p/\sigma_p)^2-1$ is a parameter related to this mean and the variance of the progressive speed distribution. The Schulz distribution (\ref{eq:schulz}) is identical to the gamma distribution, modulo a simple transformation of parameters (see section \ref{sec:simulations}).
Physically it displays the correct general features (going through the origin and peaked), and mathematically it allows us to make analytical progress. 

\subsection{Approximate ISF expressions}

We summarise here limiting expressions of the ISF (\ref{eq:IFDalgaederiv3_new}), which will be employed in section \ref{sec:results} to analyse simulations of helical swimmers and aid interpretation of experiments with real helical swimmers. Deriving these from the full ISF is straightforward, as demonstrated in the next section for the case when back-and-forth and helical swimming motions are uncoupled. 

\subsubsection{Uncoupled back-and-forth and helical swimming motions.}

The swimming of realistic swimmers, such as the biflagellate algae {\it Chlamydomonas reinhardtii} and {\it Dunaliella salina}, displays both back-and-forth and helical swimming motions. In this case, the full ISF (\ref{eq:IFDalgaederiv3_new}) is required, which is hard to simplify analytically much further (integration is straightforward over either, but not both, random phases). For helical swimmers like {\it C. reinhardtii} that move along tight helices with well-separated helical ($1/\omega_h$) and back-and-forth ($1/\omega_b$) motion timescales \cite{RufferNultsch85, Racey1981, Racey1983},  the ISF (\ref{eq:IFDalgaederiv3_new}) can be written in terms of the non-dimensional small parameters $\epsilon:=\omega_h R/v_p$ and $\nu:=\omega_h/\omega_b$, so that (see \ref{sec:appendixB})
\be\label{eq:ISFweakcouple}
\eta\approx q v_p \tau\cos\theta+ X_b \cos\left(\frac{\omega_b\tau}{2}+\phi_b\right)- X_h \sin\left(\frac{\omega_h\tau}{2}+\phi_h\right),
\ee
where we have defined 
\be\label{eq:Xb}
X_b(\theta, \tau)\equiv 2 q A_b \sin\left(\frac{\omega_b \tau}{2}\right)\cos\theta\left(1 - \epsilon \sin\phi_h \tan\theta \right)%\frac{\omega_h R}{v_p} 
\ee
and
\be\label{eq:Xh} 
X_h(\theta,\tau)\equiv 2 q R  \sin\left(\frac{\omega_h \tau}{2}\right) \sin\theta,
\ee
and we have used the trigonometric identity $\cos(A)-\cos(B)=-2\sin[(A+B)/2] \sin[(A-B)/2]$ and similarly for $\sin(A)-\sin(B)$. Substituting equation (\ref{eq:ISFweakcouple}) into the ISF (\ref{eq:IFDalgaederiv3_new}), we can integrate over the random phase $\phi_b$ to obtain:
\bea\label{eq:ISFnobreast}
&&f(q,\tau) \approx\frac{1}{2}  \int_0^{\pi}\sin \theta  d\theta \int_{0}^\infty   P_s(v_p) e^{i q v_p \tau\cos\theta}d v_p  \nonumber\\
&& \times\int_0^{2\pi}e^{i X_h \sin\left(\frac{\omega_h\tau}{2}+\phi_h\right)} J_0(X_b) d\phi_h
\eea
where $J_0(X_b) = (1/2\pi) \int_{0}^{2\pi}e^{iX_b \cos\left(\frac{\omega_h\tau}{2}+\phi_b\right)}d\phi_b$ is the zeroth order Bessel function of the first kind \cite{AbramowitzStegun}. As shown in \ref{sec:appendixB}, if $q  A_b\epsilon$ is small, a condition always met at large scales where $q A_b\ll1$, it is possible to further integrate the ISF over $\phi_h$. For swimmers with a Schulz distribution (\ref{eq:schulz}), we can also integrate over swimming speeds $v_p$ to obtain the ISF
\bea\label{eq:helical_schulz_bstrokeandhelix}
&&f(q,\tau)\approx \int_{0}^{1}  W(\chi) J_0\left[2 q A_b \sin\left(\frac{\omega_b \tau}{2}\right) \chi\right]  \nonumber\\
&& \times J_0\left[2 q R \sin\left(\frac{\omega_h \tau}{2}\right)\sqrt{1-\chi^2}\right]  d\chi.\eea
where we have changed variables to $\chi=\cos\theta$, and defined a `ballistic kernel' function
\be\label{eq:ballistic_kernel}
W(\chi)\equiv \frac{\cos\left[(Z+1) \tan^{-1}\left(\Lambda\chi\right)\right]}{\left[1+\left(\Lambda\chi\right)^2\right]^{(Z+1)/2}}.
\ee
We have given it this name because integrating over $\chi$ provides the ballistic model \cite{Martinezetal12}, see Equation (\ref{eq:helical_schulzbal}). We have also used definitions (\ref{eq:Xb}, \ref{eq:Xh}) to make swimming parameter dependencies evident and defined the constant $\Lambda=q \overline{v}_p \tau/(Z+1)$, where we recall $Z=(\overline{v}_p/\sigma_p)^2-1$ is a parameter from which the standard deviation $\sigma_p$ of the progressive speed distribution can be obtained.

\subsubsection{Pure helical swimming (negligible breastroke motion).} Some microswimmers, such as dinoflagellates \cite{Shengetal07}, are known to have negligible or nonexistent oscillatory back-and-forth motion ($A_b\approx 0$ for all $q$ values). In this case, since $J_0(0)=1$, the ISF (\ref{eq:helical_schulz_bstrokeandhelix}) simplifies to
\be\label{eq:helical_schulz}
%f(q,\tau)\approx \frac{1}{2}\int_{-1}^{1}   W(\chi) J_0\left[2 q R \sin\left(\frac{\omega_h \tau}{2}\right)\sqrt{1-\chi^2}\right]  d\chi
f(q,\tau)\approx \int_{0}^{1}   W(\chi) J_0\left[2 q R \sin\left(\frac{\omega_h \tau}{2}\right)\sqrt{1-\chi^2}\right]  d\chi,
\ee
where all quantities have been defined above.  In what follows, to understand the effect of helical swimming on the ISF, we also consider swimmers with speed distribution $P_s(v_p)=\delta(v_p-v^1_p)$, where $v^1_p$ is the single progressive speed possessed by all swimmers. In this case, the ISF is still given by (\ref{eq:helical_schulz}), but with the substitution $W \to W_1$, where 
 \be\label{eq:1speed_ballistic_kernel}
W_1(\chi) \equiv \cos (q v^1_p \tau\chi).
 \ee
is the single speed ballistic kernel.

\subsubsection{Pure back-and-forth swimming (negligible helical motion).} It is also of interest to consider the case of negligible helical motion ($R\approx 0$ for all $q$). In this case the ISF (\ref{eq:helical_schulz_bstrokeandhelix}) reduces to
\be\label{eq:helical_schulz_nohelix}
%f(q,\tau)\approx \frac{1}{2}\int_{-1}^{1}  W(\chi) J_0\left[2 q A_b \sin\left(\frac{\omega_b \tau}{2}\right) \chi\right] d\chi,
f(q,\tau)\approx \int_{0}^{1}  W(\chi) J_0\left[2 q A_b \sin\left(\frac{\omega_b \tau}{2}\right) \chi\right] d\chi,
\ee
where $W$ is given by equation (\ref{eq:ballistic_kernel}). However, we note that, in the absence of helical motion, $\Lambda=q \overline{v} \tau/(Z+1)$, where $\overline{v}$ is the average swimming speed and $Z=(\overline{v}/\sigma)^2-1$ provides the standard deviation $\sigma$. The distinction between progressive and along-helix speeds is not relevant for this model: there is only one speed, which we denote by $v$.
\subsubsection{Ballistic swimming (negligible back-and-forth and helical motion).} Finally, we consider the limit of ballistic motion, where both back-and-forth and helical motion are neglected. It is then possible to integrate the ballistic kernel to obtain
\be\label{eq:helical_schulzbal}
f(q,\tau)\approx \int_0^{1}  W(\chi) d\chi = \frac{1}{Z\Lambda} \frac{\sin(Z \tan ^{-1}\Lambda )}{(1+\Lambda ^2)^{Z/2}}.
\ee
This ISF, just as the one for pure breastroke motion (\ref{eq:helical_schulz_nohelix}), provides the mean speed $\overline{v}$ and standard deviation $\sigma$. When applied to particle dynamics that are not purely ballistic, Eq. (\ref{eq:helical_schulzbal}) yields effective quantities, that are scale-dependent, as we shall see later, and has been previously used to model the swimming of bacteria \cite{Wilsonetal11} and algae \cite{Martinezetal12}. 

For swimmers with a single speed, as for the pure helical swimming case, we can make the substitution $W \to W_1$, where $W_1$ is given by equation (\ref{eq:1speed_ballistic_kernel}) but with a speed $v^1$ replacing the progressive one. Then equation (\ref{eq:helical_schulzbal}) integrates to
\be\label{eq:1speed_helical_schulzbal}
f(q,\tau)\approx \sinc(q v^1 \tau).
\ee

\section{Numerical methods and DDM analysis}

We performed simulations of swimmers undergoing helical and/or back-and-forth motions (in 3D). These were carried out to demonstrate that it is possible to recover swimming parameters from the simulated ISFs using DDM analysis and the ISF models just presented. ISFs were obtained by applying such analysis to `microscopy-like' pseudo-image sequences derived from the coordinates of simulated swimmers, as detailed below. 

\subsection{Simulations and pseudo-image sequence generation \label{sec:simulations}}

%We perform computer simulations to validate our theory and identify limits of extracting useful info from the ISF... Particle trajectories are generated and time-lapse "microscopy-like" images recorded and subjected to DDM analysis.

We use an individual based model, adapted from \cite{Bearon13}, to simulate the dynamics of swimmers with both helical and back-and-forth motions. Each swimmer $j$ obeys the dynamical system
\be\label{eq:helicalsim}
\dot {\mathbf{r}}_j= v_{h,j}(t) \mathbf{p}_j,\mbox{~~~~~} \dot{\mathbf{p}}_j  = \omega_h \mathbf{n}_{0\,j} \times \mathbf{p}_j,
\ee
%\bea\label{eq:helicalsimr2}
%&&\dot {\mathbf{r}_i}= v_i \mathbf{p}_i\\
%;\mbox{~~~~~} 
%&&\dot{\mathbf{p}_i}  = \omega_{\rm H} \mathbf{n}_{0\,i} \times \mathbf{p}_i \label{eq:helicalsimp2}
%\eea
%
where $\mathbf{r}_j$ is the position of the centre of oscillation of the $j$-th swimmer and $\mathbf{p}_j= \mathbf{v}_{h,j}/v_{h,j}$ its swimming direction, which rotates around the direction $\mathbf{n}_{0,j}$ (the helical axis direction for swimmer $j$, see figure 1). Dots denote time derivatives. The speed along the helix is given by
 %\green{Adding an illustration for $\mathbf{p}$, $\mathbf{n}$ and $\gamma$ to figure 1 would help here}); 
%
\be\label{eq:helicalsim_speed}
v_{h,j}(t)= v_{h,j} + \omega_b A_b \sin{\omega_b t},
\ee
a superposition of a linear translation with constant speed $v_{h,j}$ and a sinusoidal oscillation that models back-and-forth motion, e.g. for swimming algae. This is analogous to what we did for the derivation of the ISF model. Using Cartesian coordinates for $\mathbf{r}_j$ and sphericals for $\mathbf{p}_j$, equation (\ref{eq:helicalsim}) is expanded in the system of component ordinary differential equations, and these were solved numerically with MATLAB using swimming parameters realistic for algae such as {\it C. reinhardtii} (see \ref{sec:appendixC} for details). As in the ISF model, all of these parameters were assumed to have a single value, with the exception of the swimming speeds. These were obtained from the distribution of speeds projected along the helix, $v_{p,j}$, provided by the Gamma distribution
\be\label{eq:gammadist}
P(v_{p,j})=\frac{1}{\Gamma(\alpha)}\frac{1}{\beta^\alpha}v_{p,j}^{\alpha-1}e^{-v_{p,j}/\beta}, 
\ee
where $\alpha=Z+1$ and $\beta=\overline{v}_{p,j}/(Z+1)$ are the distribution parameters (substituting these into (\ref{eq:gammadist}) yields the Schulz distribution (\ref{eq:schulz})). Using relation (\ref{eq:helicalspeed}) we then obtain the speed along the helix $v_{h,j}=\sqrt{v_{p,j}^2+ (\omega_h R)^2}$, which can be substituted into (\ref{eq:helicalsim}) for each swimmer. The swimming direction, $\mathbf{p}_j$, and beat phase are initialised randomly (again, see \ref{sec:appendixC} for details), while the initial values for the helix angle $\gamma_j$ between $\mathbf{p}_j$ and $\mathbf{n}_{0,j}$ can be obtained from the initial along-helix, $v_{h,j}$, and progressive, $v_{p,j}$, speeds. This is given by $\mathbf{p}_{j}\cdot\mathbf{n}_{0,j}=v_{p,j}/v_{h,j}=\cos \gamma_j$, and needs to be prescribed for each simulated swimmer, as it is a function of speed. Image sequences of the simulated helical swimmers were constructed from the simulations by generating a Gaussian pseudo-diffraction spot for each point position $(x_i,y_i,z_i)$, as detailed in \ref{sec:appendixC}. 

\subsection{Differential dynamic microscopy analysis}

For each simulated movie, we calculate the differential image correlation functions (DICF), $g(\vec{q},\tau)$, i.e. the power-spectrum of the difference between pairs of images delayed by time $\tau$. Under suitable conditions, such as isotropic motion, $g(q,\tau)=\left<g(\vec{q},\tau)\right>_{\vec{q}}$ is related to the ISF $f(q,\tau)$, the $q^{\mathrm{th}}$ mode of the number density autocorrelation function, also called the normalised dynamic structure factor:
\be\label{eq:DICF}
g(q,\tau)=A(q)\left[1-f(q,\tau)\right]+B(q),
\ee
with $B(q)$ the instrumental noise and $A(q)$ the signal amplitude. Details can be found elsewhere \cite{Martinezetal12}. For a given simulated movie, the $g(q,\tau)$s were fitted as a function of $\tau$ either for each $q$ independently or simultaneously over a given $q$ range. A range of models were used. For simplified models that ignore the helical path, we found that applying a fitting weight at longer times allowed better recovery of the input parameters. Details are mentioned in each figure caption when relevant. For practicality, we present directly ISFs extracted from equation (\ref{eq:DICF}) using the fitted parameters $A(q)$ and $B(q)$. 

\section{Results\label{sec:results}}

\subsection{Pure helical swimming}

\begin{figure}[tbh!]
   \begin{center}
      \includegraphics[width=0.8\linewidth]{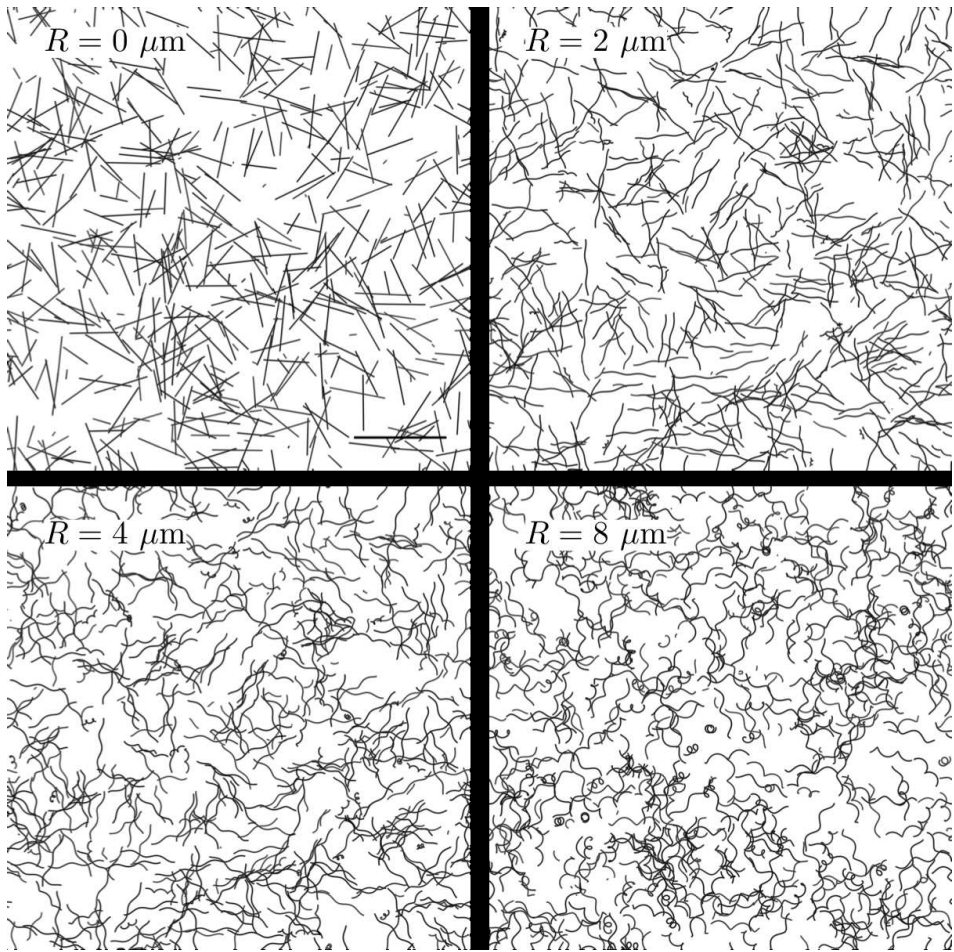}
      \caption{Examples of helical trajectories obtained from simulated movies with single progressive speed $v_p^1=120~\mu$m/s, $f_h=2$~Hz, and R=0, 2, 4, and 8~$\mu$m as indicated. Each image were obtained by accumulating images over 1s from movies  (with $L_z=1000~\mu$m, see Appendix C). Scale bar corresponds to 200 $\mu$m.}
      \label{fig:trajectories}
   \end{center}
\end{figure}

\begin{figure}[tbh!]
   \begin{center}
      \includegraphics[width=0.8\linewidth]{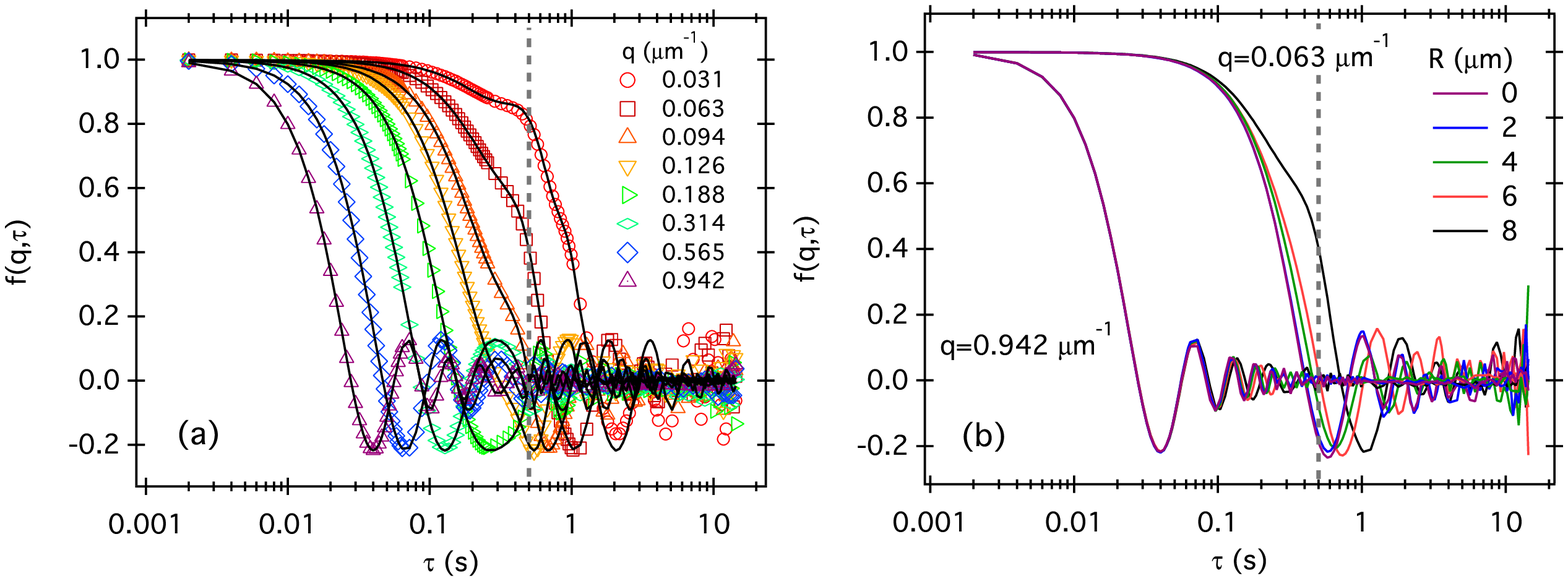}
      \includegraphics[width=0.8\linewidth]{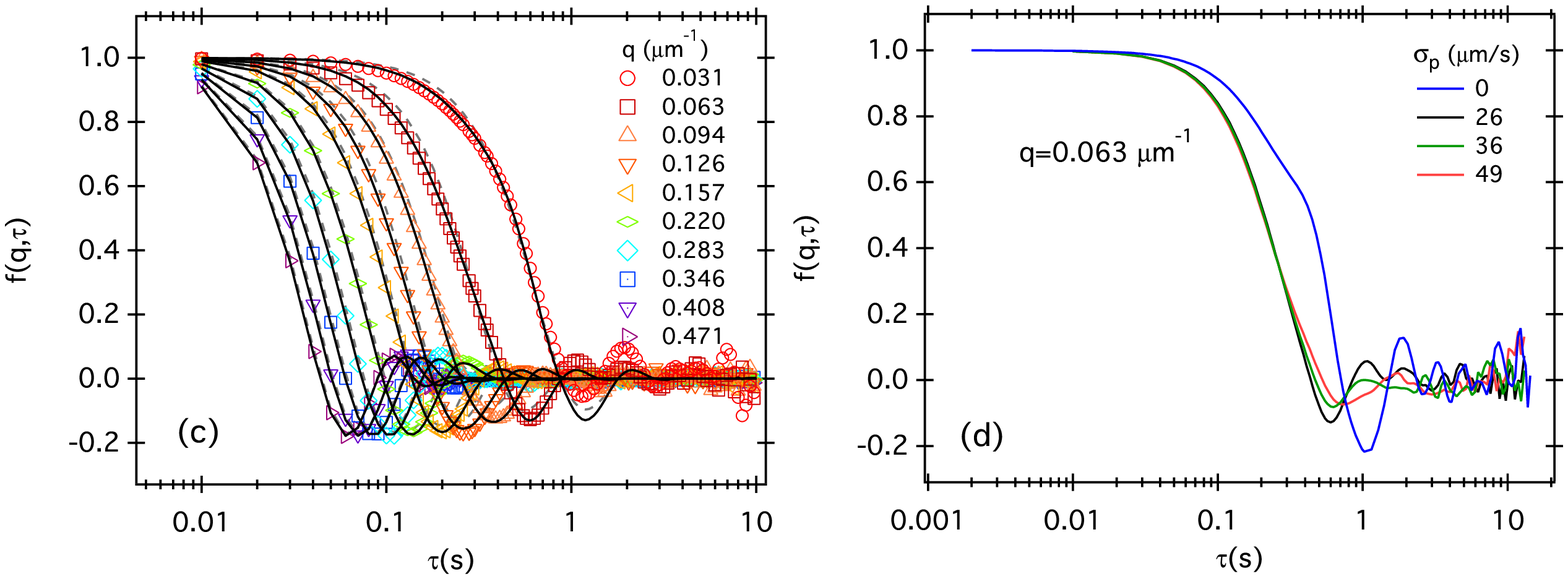}
      \caption{Pure helical swimming ISFs as a function of delay $\tau$ for different $q$ values: (a) from simulations of helical swimmers with a single speed (input parameters: single along-the-helix $v^1_h=120$ $\mu$m/s, helical frequency $f_h=2$ Hz and radius $R=8$ $\mu$m). Thick lines are independent fits to the ISFs for each $q$ using the single speed helical ISF model (\ref{eq:helical_schulz}), as discussed in the text; (b) as in (a), but showing the effect of varying helical radius $R$ at small $q$ (large length-scales) and large $q$ (small length-scales). A \textit{kink} is highlighted in the ISF (vertical dotted lines) at small $q$, which is a characteristic of helical motion that vanishes as $R$ is reduced to $0$ (ballistic motion); (c) for swimmers with a Schulz speed distribution $P_s(v)$ with mean progressive speed $\overline{v}_p=120$ $\mu$m/s and standard deviation $\sigma_p\approx26.2$ $\mu$m/s (helical parameters as above). Thick lines indicate a global fit over all $q$ using the helical model (\ref{eq:helical_schulz}), while dashed lines denote independent fits to the ISF for each $q$ using the ballistic model (\ref{eq:1speed_helical_schulzbal}); (d) As in (c), but showing the effect of increasing $\sigma_p$ at fixed (small) $q$. The signature helical \textit{kink} is no longer obvious, but helical swimming parameters can nevertheless be extracted from the data by a global fit using the helical model (\ref{eq:helical_schulz}), see text.}
      \label{fig:ISF_helical}
   \end{center}
\end{figure}

We start by analysing simulations of swimmers undergoing pure helical swimming (no back-and-forth motion). To avoid the complications of swimming speed heterogeneity, we consider first a population of swimmers swimming with a single speed. Typical trajectories are shown in Figure \ref{fig:trajectories}.
%Representative simulated trajectories for this case are shown in Figure ADD FIGURE (see also movie ADD MOVIE in Supporting Information).
Figure \ref{fig:ISF_helical}a shows the ISFs as a function of delay time $\tau$ at several values of $q$ for a population of swimmers with a single along-the-helix speed $v^1_h=120$ $\mu$m/s, helical radius $R=2$ $\mu$m, and helical frequency $f_h=2$ Hz. The decay is qualitatively different depending on the value of $q$, which defined the length scale of interest. Indeed, equations (\ref{eq:helical_schulz}-\ref{eq:1speed_ballistic_kernel}) suggest that the characteristic times of ballistic and helical contribution scale as $\tau_b \sim 2\pi/q v_p$ and $\tau_h\sim2/f_h$, respectively, so that there exists a crossover $q_c\sim\pi f_h/v_p \approx 0.1$ $\mu$m$^{-1}$ for which $\tau_b=\tau_h$. Thus, helical motion should provide the fastest contribution to the dynamics for $q\ll q_c$, and the slowest at $q\gg q_c$.

For small $q$, i.e. $q\ll q_c$ (large length-scales), the ISF displays two main decays, corresponding to ballistic and helical contributions to the swimming motion. In particular, a kink in the ISF can be seen at time $\tau\simeq 0.5$ s, independent of $q$, for $q\lesssim0.1$ (see vertical dashed line in figure \ref{fig:ISF_helical}a), which corresponds to the characteristic period $1/f_h=0.5$ s of helical swimmers rotating at $f_h=2$ Hz (the simulation input frequency). At large values of $q$, i.e. $q\gg q_c$, the ISF shows a single decay, which fully decorrelates on time-scales $\tau\ll \tau_h$. This is characteristic of ballistic motion: large $q$ values correspond to small length scales, where helical trajectories are ballistic to a good approximation, so helical features in the ISF are not pronounced (or easily distinguished). Additionally, the amplitude of the helical process scales with $q$ and thus decreases with $q$ as expected from the $J_0$ term in equation (\ref{eq:helical_schulz}). 

The effect of helical rotation on the ISF, via the observation of a kink, is strongly sensitive to the helical swimming parameters. For example, the characteristic kink evident for single speed swimmers (figure~\ref{fig:ISF_helical}(a)) is less apparent in the following instances when the radius $R$ of the helical path is decreased (figure~\ref{fig:ISF_helical}b), or when swimmers obey a speed distribution with non-zero standard deviation (figure~\ref{fig:ISF_helical} (c) and (d)).

The signature of helical motion is nevertheless encoded in the swimmer ISF. We demonstrate in the next sections that fitting the ISFs with appropriate models derived in section \ref{sec:theory} allows to measure the 3D motility parameters, with good accuracy, from 2D movies of the swimmers. 

%
%%%% new FIGURE 2: trajectories of HELICAL microswimmers %%%%%%%%%

\subsubsection{Motility parameters from the simulated swimmer ISF: single speed swimmers.}

\begin{figure}[tbh!]
   \begin{center}
      \includegraphics[width=1\linewidth]{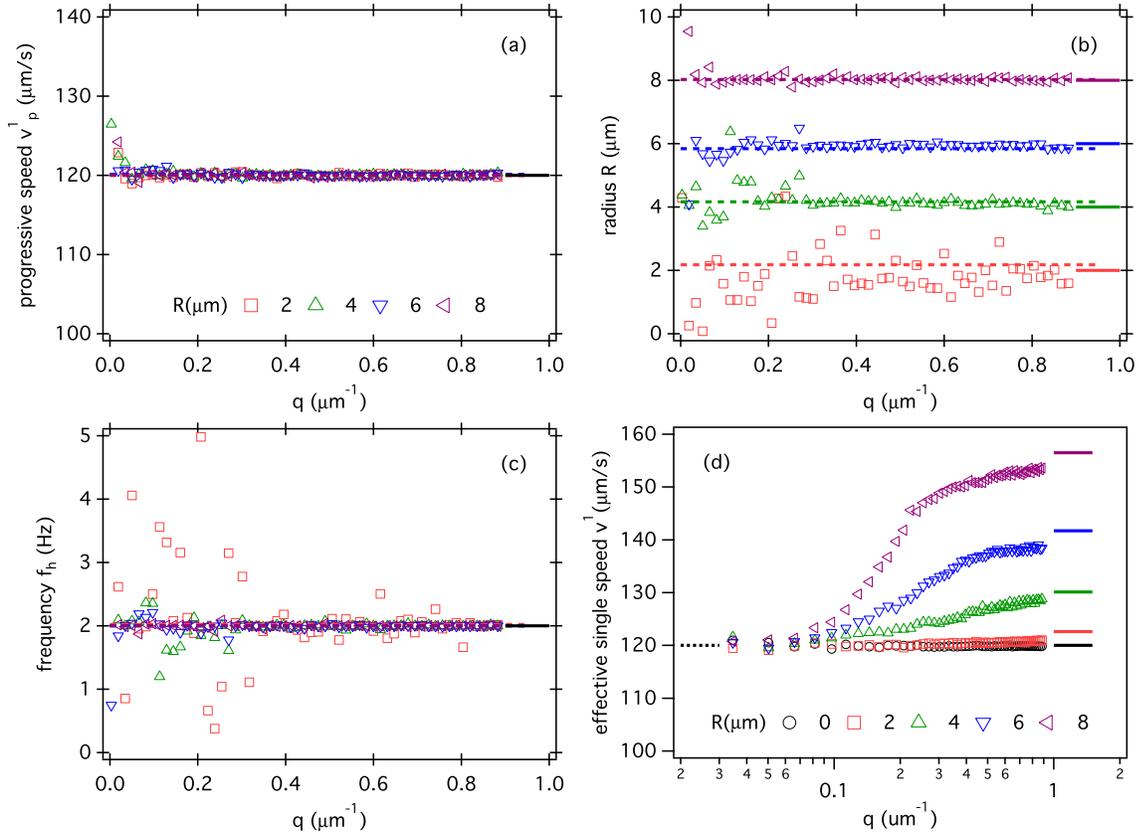}
      \caption{Swimming parameters from fits to ISFs generated from single speed helical swimmer simulations (inputs: single speed $v^1_p=120$ $\mu$m/s, helical frequency $f_h=2$ Hz, helix radius $R$ values in legends). Measured parameters are shown as function of $q$ for different $R$. Panels (a)-(c) show swimming speed $v^1_p$, helical radius $R$ and frequency $f_h$, respectively. Symbols are results obtained by independent fits of the ISF for each $q$ using the single speed helical model (\ref{eq:helical_schulz}). Dotted lines are results from global fits of the ISFs over the range $0.05\lesssim q \lesssim 0.45$ $\mu$m$^{-1}$. Solid lines are input values used for the simulating the swimmers trajectories. (d) effective swimming speed $v_1$ obtained by fitting the same ISFs with a simpler ballistic model using equation (\ref{eq:helical_schulzbal}). At high $q$ (small length scales), $v^1\rightarrow v^1_h$, matching the input values (thick lines) to $\lesssim 3\%$. At low $q$ (large length scales), the speed $v^1$ equals the input value of the progressive speed $v_p$ (dotted line). Note: here and in subsequent figures, data errorbars are smaller than the pointsize. 
      }
      \label{fig:vq_ballistic}
   \end{center}
\end{figure}

We first analyse the ISF of a population of helical swimmers with one swimming speed. The ISF is fitted using equation (\ref{eq:helical_schulz}) with $W$ given by equation (\ref{eq:1speed_ballistic_kernel}). Fitting for each $q$ independently, we can infer the swimming parameters as a function of $q$. Figure \ref{fig:vq_ballistic} displays the $q$ dependence of the progressive single swimming speed $v_p^1$, helix radius $R$ and frequency $f_h$ (panels a-c respectively). We see that parameters are recovered well with better results at large $R$. For small $R$, features of helical motion in the ISF are not pronounced (e.g. see figure \ref{fig:ISF_helical}(b) and (c) for $R=2$ $\mu$m), resulting in less accurate parameter recovery. However, global fits over given ranges of $q$ permit accurate recovery of input values, see dotted lines in figure \ref{fig:vq_ballistic} and figure caption.

It is also interesting to analyse the ISF of helical swimming with the simpler ballistic model (\ref{eq:1speed_helical_schulzbal}), which does not account for helical swimming motions. %This, as we will show in section XXX, allows to measure 3D motility statistics based on simple 2D tracking information.
The ballistic model yields a speed that varies with $q$ for non-zero $R$ (figure \ref{fig:vq_ballistic}(d)). The ballistic model does not account for helical motion and thus provides an effective speed, with a meaning that depends on length scale. For small length scales (high $q$), it represents the along-helix speed $v^1_h$, matching the input values to $\leq 3\%$, where the effects of helical motion are not appreciable. At large length scales ($q\ll q_c$), the speed tends towards the progressive speed which, rearranging equation (\ref{eq:helicalspeed}), is given by $v^1_p=\sqrt{(v^1_h)^2-(\omega_hR)^2}$. We recall this is the projection of the along-helix speed onto the helix axis: at large length scales, helical motion is averaged out over several helical pitches.
%
%%%% FIGURE 3: HELICAL PARAMETERS SINGLE SPEED %%%%%%%%%%%
\subsubsection{Motility parameters from the simulated swimmer ISF: Schulz speed distribution.}
\begin{figure}[tbh!]
   \begin{center}
      \includegraphics[width=1\linewidth]{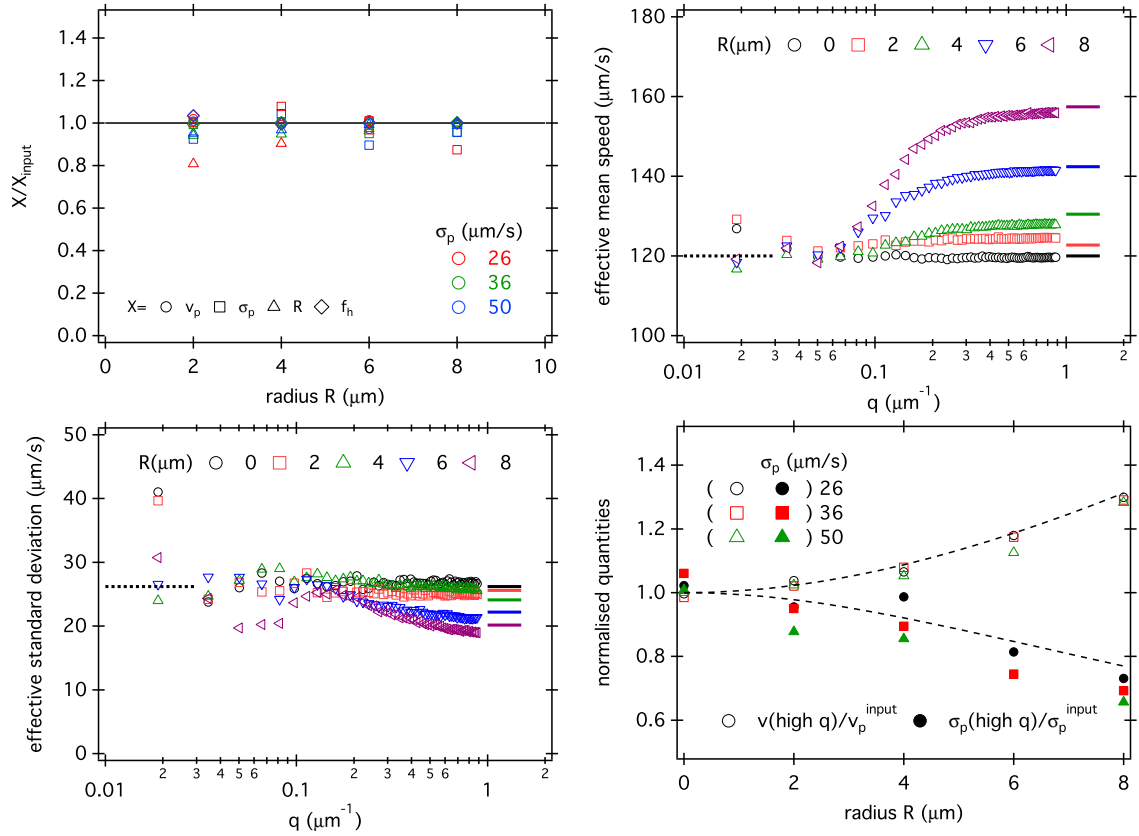}
      \caption{Swimming parameters from fits to ISFs obtained from simulations of helical swimmers with a Schulz speed distribution (input values: $\overline{v}_p=120\mu$m/s and $f_h=2$ Hz, $\sigma_p$ and $R$ in legends). (a) Dependence on radius $R$ of the observed to input swimming parameters (see legend) obtained from a global fit of the ISFs, over the range $0.05\lesssim q \lesssim 0.45\mu$m$^{-1}$, using the helical model for swimmers with a Schulz distribution using equation (\ref{eq:helical_schulz}) and a fitting weight at longer times. (b) mean speed $\bar{v}$ and (c) standard deviation $\sigma$ of the effective speed distribution as a function of $q$ extracted by fitting the ISFs with the ballistic model using equation (\ref{eq:helical_schulzbal}). Thick lines denote the input values of mean along-helix speed $\overline{v_h}$ and standard deviation $\sigma_h$ obtained from input values of the progressive speed $\overline{v_p}$ and standard deviation $\sigma_p$ (dotted lines) using equations (\ref{eq:avg_inst_speedmom}) and (\ref{eq:inst_speedvar}), respectively. (d) Ratio of high $q$ (small length scales) mean speed $\bar{v}$ (open symbols) and standard deviation $\sigma$ (filled symbols) obtained by fitting the ballistic ISF (\ref{eq:helical_schulzbal}), to simulation input values as a function of $R$. The normalised speed and standard deviation increases and decreases, respectively, monotonically with $R$, reflecting the dependence of the mean along-helix swimming speed $\overline{v_h}$ and standard deviation $\sigma_h$ on $R$ given by equations (\ref{eq:avg_inst_speedmom}) and (\ref{eq:inst_speedvar}) and plotted as dashed lines. 
      }
       \label{fig:vq_ballisticPv}
   \end{center}
\end{figure}

We show in figure~\ref{fig:ISF_helical} typical ISFs for helical swimmers with a Schulz swimming speed distribution. As mentioned, a distribution of speeds makes helical features in the ISF more subtle and they qualitatively look similar to swimmers with pure ballistic motion ($R=0$) (figure~\ref{fig:ISF_helical}(d)). Here, we fit the ISFs with a model for helical swimmers with a Schulz swimming speed distribution given by equation (\ref{eq:schulz}), using (\ref{eq:helical_schulz}) for the ISF with $W$ given by the ballistic kernel (\ref{eq:ballistic_kernel}). For smaller values of the helix radii, we found that global fitting the ISFs provides more reliable results than fitting each $q$ independently (data not shown). 
%As mentioned, a distribution of speeds makes helical features in the ISF more subtle, and also harder to fit. We thus carried out global fits over all $q$, which provides more reliable results than fitting each $q$ independently.
Figure~\ref{fig:vq_ballisticPv}(a) shows the ratio of the swimming parameters extracted from the global fit to simulation input values, as a function of the helix radius $R$. The parameters are denoted collectively as $X=\overline{v}_p, \sigma, R, f_h$, the swimming speed, variance, helix radius and frequency, respectively (corresponding input values are denoted by $X_{input}$). It is clear from figure~\ref{fig:vq_ballisticPv}(a) that most parameters are recovered with good accuracy (within $\approx10\%$) for all $R$. This demonstrates the applicability of the helical ISF model for swimmers with a Schulz distribution and shows how this model is sensitive to helical signatures not evident from inspection of the ISF (see figure~\ref{fig:ISF_helical}(b)). At lower values of $R$, while $\overline{v}_p$ and $f_h$ are recovered to within a few percent across the range, the accuracy of recovery for $R$ is less accurate. This is because, at such small $R$ values, the helical signature in the ISF is too weak for even the global fit to pick up. 

It is useful to also analyse the Schulz swimmer ISF using the ballistic model, as we did for single speed swimmers. In this case we can fit the ISF with equation (\ref{eq:helical_schulzbal}) independently for each value of $q$. The results for the effective mean swimming speed $\overline{v}$ and distribution standard deviation $\sigma$ are shown in figure \ref{fig:vq_ballisticPv}(b) and (c), respectively, as a function of $q$ for different values of $R$.
%We note that the same mean progressive speed of $120\,\mu$m/s has been used as input in all the simulations considered here.
As for the single speed case, fitting the ballistic model to the helical swimmer ISF results in a $q$-dependent speed $\overline{v}(q)$, whose value increases from its progressive value $\overline{v}_p$ at low $q$ to its along-helix value $\overline{v}_h$ at high $q$. A similar trend is observed for $\sigma$ although the data can be noisier.

As is clear from Figure \ref{fig:vq_ballisticPv}b, the mean along-helix speed value grows with the helix radius $R$. The functional form for this variation, provided by equation (\ref{eq:avg_inst_speedmom}), is explicitly plotted in Figure \ref{fig:vq_ballisticPv}d and compared with the ratio between the high $q$ value of the mean speed from fits and the simulation input. Similarly, the theoretical and fitted ratios of the corresponding standard deviations, using equation (\ref{eq:inst_speedvar}) are plotted on the same graph. The speed and standard deviation data follow the expected variation with $R$, which provides a useful check that the dependencies predicted by our analytical theory are borne out by the simulated experiments. 
%
%%%% FIGURE 4: HELICAL PARAMETERS P(v) %%%%%%%%%%%

\subsection{Pure back-and-forth swimming\label{sec:purebreastroke}}

\begin{figure}[tbh!]
   \begin{center}
            \includegraphics[width=1\linewidth]{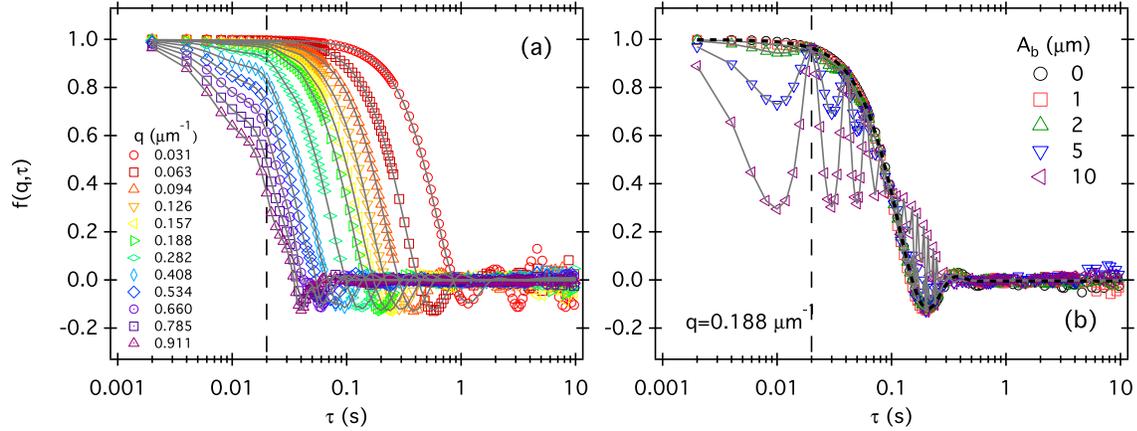}
      \caption{Back-and-forth swimming ISFs as a function of delay $\tau$: (a) for several $q$ values (see legend) from simulations of back-and-forth swimmers with a Schulz speed distribution (input parameters: mean progressive speed $\overline{v_p}=120$ $\mu$m/s, standard deviation $\sigma_p=26.2$ $\mu$m/s,
%      $\sigma/v=0.22$, 
      back-and-forth frequency $f_b=50$ Hz and amplitude $A_b=2$ $\mu$m); (b) for different values of the amplitude $A_b$. The characteristic, $q$-independent signature of the back-and-forth oscillation is a dip in the ISF at the back-and-forth timescale $1/f_b=0.02$ s indicated as vertical dashed lines. Increasing $A_b$ makes this feature more pronounced. Continuous grey lines in (a $\&$ b) represent fits for each $q$ using the back-and-forth model with equation (\ref{eq:helical_schulz_nohelix}) for the ISF. In (b) the dotted line is a fit to $A_b=0$ using ballistic model.
}
      \label{fig:ISF_bstroke}
   \end{center}
\end{figure}

Next, we analyse simulations of swimmers with only back-and-forth motion (no helical swimming) and a Schulz distribution of swimming speeds using typical swimming parameters for biflagellate microalgae, like {\it Chlamydomonas spp.} or {\it Dunaliella spp.}. The ISF obtained from these simulations is shown for different values of $q$ in figure \ref{fig:ISF_bstroke}(a). As for helical swimming, a $q$-independent signature of back-and-forth motion is evident. This is oscillatory in time and its amplitude decreases with decreasing back-and-forth amplitude $A_b$, vanishing for $A_b=0$, see figure \ref{fig:ISF_bstroke}(b), as expected from the $J_0$ term in equation (\ref{eq:helical_schulz_nohelix}). The short-time decay of the ISF ($\tau\lesssim0.02$ s) corresponds to the back-and-forth swimming oscillation, while longer time decay corresponds to ballistic motion. Subsequent oscillations at longer times in the decay, before the noise floor is reached, reflect oscillations in the Bessel function $J_0$ (see equation (\ref{eq:helical_schulz_nohelix})). In contrast to the helical case, the back-and-forth signature in the ISF is more evident at high $q$ values, because of its small length-scale amplitude.
%
%%%% FIGURE 5: B&F SWIM P(v) %%%%%%%%%%%

\subsubsection{Motility parameters from the simulated swimmer ISF.} 
\begin{figure}[tbh!]
   \begin{center}
                        \includegraphics[width=1\linewidth]{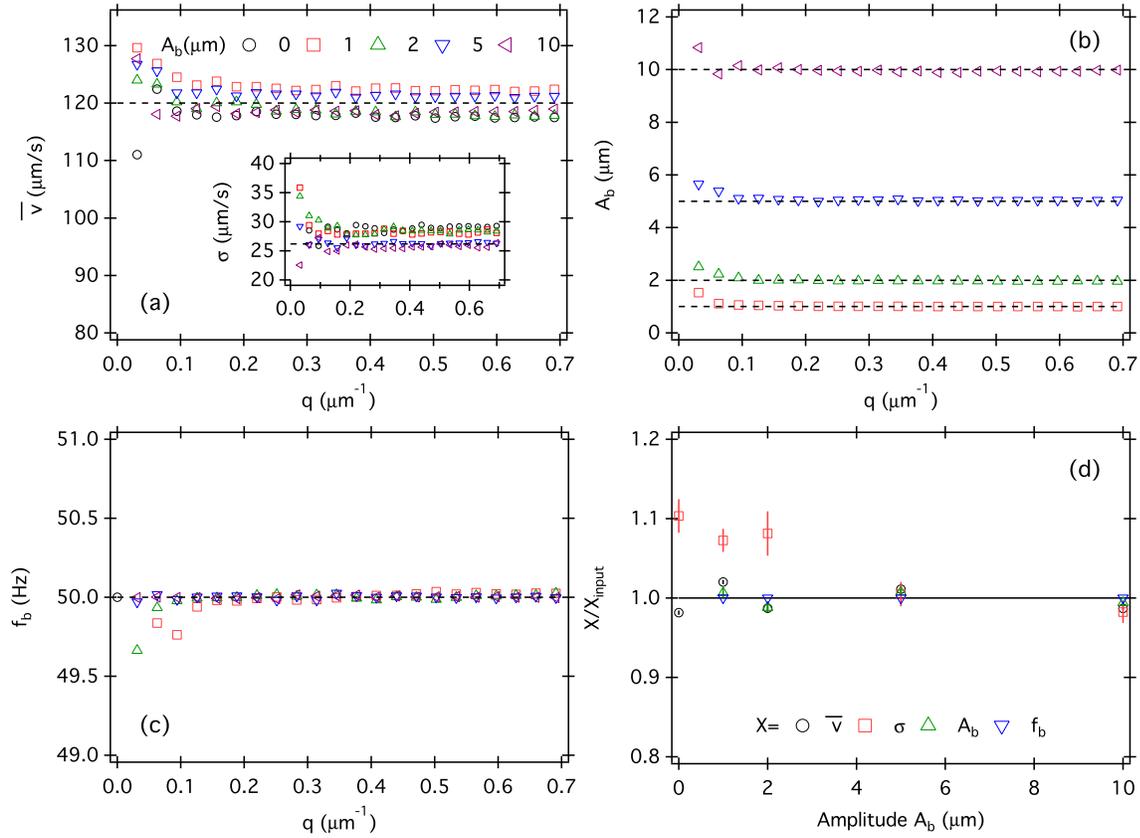}
      \caption{Swimming parameters from fits to ISFs from simulations of back-and-forth swimmers with a Schulz speed distribution (input values: $\overline{v}=120\mu$m/s, $\sigma=26.2\mu$m/s and $f_b=50$ Hz; amplitudes $A_b$ are as indicated in the plots). (a) mean speed, (b) beating amplitude $A_b$, and (c) beating frequency $f_b$ obtained by fitting the ISFs with the back-and-forth model using equation (\ref{eq:helical_schulz_nohelix}). Dashed lines are input values of the simulations. (d) Ratio of observed to input swimming parameters (as shown in legend) as a function of $A_b$.
      }
      \label{ffig:ISF_bstrokeparams}
   \end{center}
\end{figure}

Analogously to the helical case, fitting the simulated back-and-forth swimmer ISF with equation (\ref{eq:helical_schulz_nohelix}) provides values for the swimming parameters. Figure~\ref{ffig:ISF_bstrokeparams}(a)-(c) shows the variation with $q$ of the mean swimming speed $\overline{v}$ and standard deviation $\sigma$ (inset), back-and-forth amplitude $A_b$ and frequency $f_b$, respectively. For small values of $A_b$, the values of $\overline{v}$ and $\sigma$ deviate from the simulation inputs at low $q$: here, the back-and-forth signal in the ISF for small $A_b$ is weak, and thus harder to fit (similarly to the small $R$ case for helical swimmers). Unsurprisingly, on the other hand, since $f_b$ and $A_b$ are $q$-independent, their values can be accurately determined across the $q$ range.  The effect of the magnitude of $A_b$ on parameter accuracy is shown in figure \ref{ffig:ISF_bstrokeparams}(d), where we plot the ratio of the high $q$ value of the swimming parameters to the input value. Note how the effect of $A_b$ on accuracy is analogous to that of $R$ for helical swimmers. 
%
%%%% FIGURE 6 %%%%%%%%%%%

\subsection{Back-and-forth and helical swimming}

\begin{figure}[tbh!]
   \begin{center}
                        \includegraphics[width=0.75\linewidth]{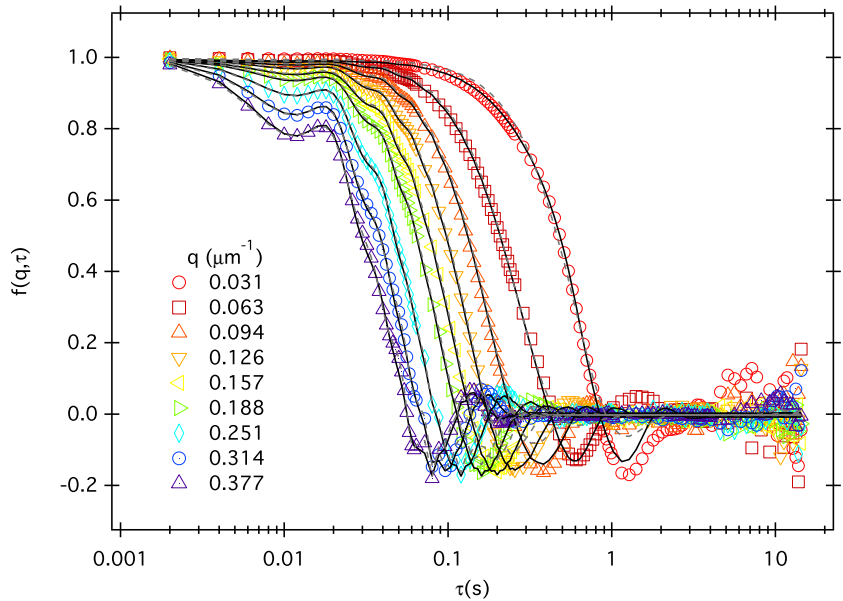}
                                             
      \caption{ISF for swimmers with back-and-forth {\it and} helical swimming motions shown for different values of $q$. Simulation input values are: $\overline{v}_p=120\mu$m/s, $\sigma_p=26.2\mu$m/s, $R=8\mu$m, $f_h=2$ Hz, $A_b=2\mu$m, and $f_b=50$ Hz. Dotted Lines are fits using equation (\ref{eq:helical_schulz_nohelix}) for the back-and-forth model. Thick lines are from global fits using equation (\ref{eq:helical_schulz_bstrokeandhelix}) for the helical and back-and-forth model.
            }
      \label{ffig:ISF_bstrokeandhelical}
   \end{center}
\end{figure}

Finally, we consider simulations of swimmers with both helical and back-and-forth motions, and a Schulz distribution of speeds. Figure~\ref{ffig:ISF_bstrokeandhelical} shows the corresponding ISF decay for different $q$ values. The ISFs encode both pure helical and back-and-forth swimming although only back-and-forth features are clearly evident.
%
%%%% FIGURE 7 %%%%%%%%%%%

\subsubsection{Motility parameters from the simulated swimmer ISF.} 
\begin{figure}
   \begin{center}
                        \includegraphics[width=1\linewidth]{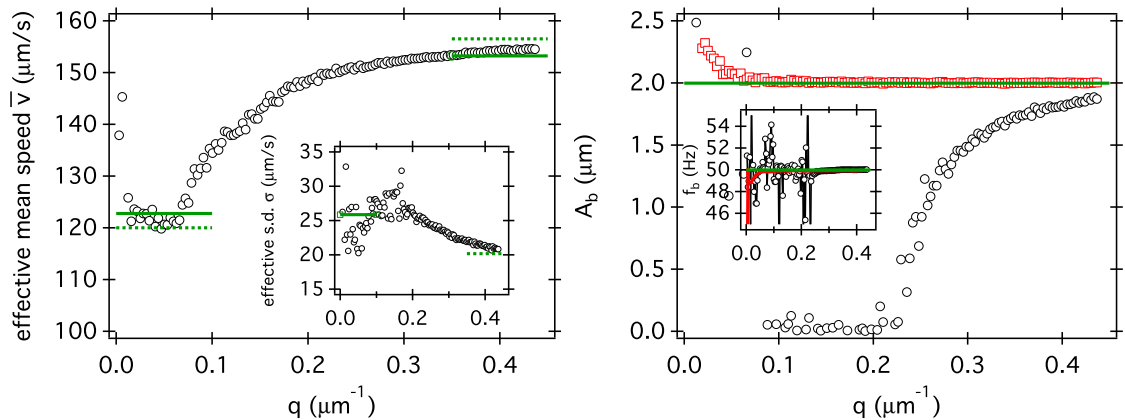}
      \caption{Swimming parameters from fits to the ISFs of Fig.~\ref{ffig:ISF_bstrokeandhelical} for swimmers with combined helical and back-and-forth motions using input parameters $\overline{v}_p=120\mu$m/s, $\sigma_p=26.2\mu$m/s, $R=8\mu$m, $f_h=2$ Hz, $A_b=2\mu$m, and $f_b=50$ Hz. Symbols are in (a) mean effective speed $\bar{v}$ and standard deviation $\sigma$ (inset); and (b) amplitude $A_b$ and frequency $f_b$ (inset) as a function of $q$ obtained by fitting independently each $f(q,\tau)$ using the back-and-forth model (no helical path) with equation (\ref{eq:helical_schulz_nohelix}). Black and red symbols correspond to fits using stronger fitting weight at longer or shorter times, respectively. Dotted lines are input values of simulations. Thick lines are fitted parameters using a global fit (no weight) with combined helical and back-and-forth motions (equation~\ref{eq:helical_schulz_bstrokeandhelix}) over the range $0.05\lesssim q \lesssim 0.45\mu$m$^{-1}$ with at low q and high q the progressive speed and the along-helix speed, respectively. Results from global fits are $\bar{v_p}=122.8\mu$m/s, $\sigma_p=25.9\mu$m/s, $R\approx7.5\mu$m/s, $f_h\approx2.0$ Hz, $A_b=2.0\mu$m, and $f_b=50.0$ Hz.
      }
      \label{fig:bstrokeandhelicalparams}
   \end{center}
\end{figure}

We fit the back-and-forth and helical swimming ISF with equation (\ref{eq:helical_schulz_bstrokeandhelix}) to obtain swimming parameters. As for the pure helical case, a global fit over a selected $q$ range provides the best accuracy, with parameter value recovery $\leq 5\%$ (figure~\ref{fig:bstrokeandhelicalparams}). %REFER TO GLOBAL FITS IN FIGURE.

 We have also fitted the ISF with the pure back-and-forth model (ignoring the helical path) given by equation (\ref{eq:helical_schulz_nohelix}), independently for each value of $q$. Figure \ref{fig:bstrokeandhelicalparams}(a) and (b) shows the mean effective swimming speed $\overline{v}$, distribution standard deviation $\sigma$, swimming amplitude $A_b$ and frequency $f_b$, respectively, as a function of $q$. We found that fitting with a stronger weight at longer times, thus ignoring short-time contributions, yielded a similar $q$ dependence we observed for the pure helical case. As in this case, this is an artefact of using a model that is unable to account for helical trajectories. As a result, $\overline{v}$ transitions from its progressive value, $\overline{v}_p$, at low $q$, to its along-helix value, $\overline{v}_h$, at high $q$, and similarly for the distribution standard deviation $\sigma$. As for the helical case, we can use equation (\ref{eq:avg_inst_speedmom}) and (\ref{eq:inst_speedvar}) to relate along-helix and progressive values for a given helical angular frequency $\omega_h$ and radius $R$. More interestingly, when the helical parameters are not known (as is the case in experiment) we can estimate the product $\omega_h R$, a measure of the rotational speed around the helical axis. Indeed, simply rearranging equation (\ref{eq:inst_speedvar}) provides:
\be\label{omegaRestimate}
\omega_h R = \sqrt{(\sigma_h^2-\sigma_p^2)+(\overline{v_h}^2-\overline{v_p}^2)}.
\ee
Thus, along-helix (high $q$) and progressive (low $q$) values of the speed and distribution standard deviation can be used to estimate $\omega_h R$  based on a simple ISF model that does not account for helical motions. From figure \ref{fig:bstrokeandhelicalparams}, we measure the following values: $v_p\approx122\mu m/s$, $\sigma_p\approx24\mu m/s$, $v_h\approx 154\mu m/s$, and $\sigma_h\approx21\mu m/s$. Thus, using equation (\ref{omegaRestimate}), we obtain $(\omega_h R)\approx 95$ $\mu$m rad/s, only $6\%$ from the simulation input value.  Finally, while oscillating components are only recovered towards higher $q$ values, fitting with a stronger weight at short-time allows determination of the oscillating component within $\leq 0.01\%$, see caption of figure \ref{fig:bstrokeandhelicalparams}.
%
%%%% FIGURE 8 %%%%%%%%%%%

\section{Discussion and Conclusions}

The experimental characterisation of the helical and oscillatory motions of microswimmers is essential for understanding their statistical mechanics \cite{Elgeti2015, Bechinger2016}, fluid mechanics \cite{GuastoRusconiStocker12, Elgeti2015, Bechinger2016}, biology \cite{Goldstein2015} and ecology \cite{Shengetal07}. In this work, we developed a theoretical framework enabling the quantification of such motions by differential dynamic microscopy (DDM), a technique allowing to infer motility statistics from videomicroscopy without the need for specialised apparatus. We modelled the intermediate scattering function (ISF) corresponding to helical, back-and-forth, and combined helical and back-and-forth motions. Individual based model simulations of swimmers with theses same motions were used to generate artificial movie sequences, from which the ISFs were extracted, as in DDM \cite{Martinezetal12} and fitted with our models. This study provides the basis for applying ensemble averaging techniques such as DDM to measure the motility statistics of microswimmers following a helical path or with a combination of helical and other motions. 

The ISFs from our simulations allow to explore characteristic ISF features emerging from helical and back-and-forth swimmer motions. In particular, for swimmers with pure back-and-forth motion, a characteristic dip in the ISF is evident at the timescale corresponding to the oscillation (figure \ref{fig:ISF_bstroke}). A similar feature is evident for pure helical swimmers, corresponding to helical rotation. This is only apparent for swimmers with a single speed, and at large enough value of the helix radius. For swimmers with a speed distribution, the helical swimming feature is not conspicuous (see figure \ref{fig:ISF_helical}), because it is spread across different length-scales for swimmers with different swimming speeds. 
%For the full helical and back-and-forth simulations, it will be interesting to compare simulated and experimental ISFs, e.g. those measured from the combined helical and breaststroke motions of the microalga {\it Chlamydomonas reinhardtii}.

Beyond predicting features to be expected experimentally, our simulated ISFs can be fitted using the models we have developed to test how accurately the simulation input parameters can be extracted using DDM. For all the models considered in this work, we find that it is possible to extract helical swimming statistics from simulated ISFs with good accuracy, even when qualitative features are not evident in the ISFs.  We note that some of the assumptions we have made in our models (e.g. that all swimming parameters but speed are delta function distributions, different swimming motions can be decoupled) may not apply in real experimental systems. More detailed models of swimmer ISFs than considered here may be formulated, e.g. when several contributions to the ISFs from different swimming motions cannot be decoupled. In this case, however, the added details (e.g. multiple integrals, additional dynamical parameters, parameter distributions) may render fitting impractical.

In alternative to detailed models, our analysis also shows that it is also possible to extract information about complex swimming motions from simple ISF models. For example, when fitting the ISF of helical and back-and-forth swimmers with a simpler model ignoring helical motion, we find that a scale-dependent effective mean speed, $\overline{v}(q)$, and a speed distribution standard deviation, $\sigma(q)$, emerge. Their low $q$ (large lengthscale) values correspond to the mean progressive speed $\overline{v}_p$ and standard deviation $\sigma_p$, while their high $q$ (small lengthscale) values represent the mean along-helix speed $\overline{v_h}$ and standard deviation $\sigma_h$. These progressive and along-helix values can be combined to estimate the product $\omega_h R$ between the helical angular frequency $\omega_h$ and radius $R$, see equation (\ref{omegaRestimate}), where $\omega_h R$ is a measure of the speed around the helical axis. Further, independent knowledge of $\omega_h$, obtainable e.g. from particle tracking, can provide an estimate of the (mean) helix radius $R$. For example, assuming a measured helical frequency $f_h=2$ Hz, we can use the value $\omega_h R\approx 95$ $\mu$m rad/s (estimated in the previous section from the data in figure \ref{fig:bstrokeandhelicalparams}) to obtain a helix radius $R\approx 7.6$ $\mu$m, which is within $5\%$ of the input value used for the simulation.

The helix radius is an intrinsically 3D property of helical swimming. Thus far, to acquire motility statistics for this parameter, specialised techniques have been used, e.g. microscopy involving multiple cameras \cite{Crenshaw2000, Boakesetal00, Drescher2009, Polinetal09, Orchard2016} or holographic microscopy \cite{Shengetal07}. Such techniques are limited in statistical accuracy and/or to relatively dilute microswimmer systems, as explained in the introduction. Gurarie {\it al} \cite{Gurarie2011} recently obtained 3D motility statistics of dilute suspensions of dinoflagellate microalgae from 2D projections of helical trajectories obtained using standard microscopy using a continuous stochastic model (CSM) of helical swimming \cite{Gurarie2011}. In this study, the dinoflagellates swam with pronounced (large radius) helical trajectories, and little other body oscillations. It would be challenging, however, to apply the CSM method to swimmers with small helical radius and back-and-forth oscillations, such as {\it C. reinhardtii}. Instead, the hybrid DDM-tracking approach proposed here has great potential to enable measurements of the helix radii for moderately concentrated suspensions of helical swimmers.  
%In standard 2D-imaging microscopy, the 2D-projection of a helix is a sinusoidal oscillation with the same amplitude as the helix because the projection encodes 3D information from helices moving in all directions. 

The efficacy of our methods and analyses needs to be tested by applying differential dynamic microscopy to suspensions of real artificial and biological swimmers. As discussed, helical swimming motions are commonplace for many types of microswimmer \cite{Bechinger2016}. Using the methods we have here developed, high quality 3D motility statistics can be collected with standard 2D-imaging microscopy. This provides economical experimental set-ups for the analysis of 3D microswimmer motility, making this important area of research accessible to researchers with a limited equipment budget, such as in schools and developing countries. The ability of our new methods to extract helical swimming parameters without the use of specialised microscopy setups could also be exploited in field work (e.g. to characterise unculturable helical swimmers, such as dinoflagellates, {\it in situ}). Future work could develop the models used in our analysis, adapting them to the large variety of interesting synthetic and biological swimmers, both currently known and to be discovered. This will allow theoretical and experimental progress in our statistical understanding of microswimmers, facilitating the development of microswimmer based biotechnologies, and aiding the characterisation of biological microswimmers in the environment. 

\section{Data availability.} The raw data (i.e. simulated movies) presented in this publication is available on the Edinburgh DataShare \cite{data_repository}.

\section{Acknowledgements}
We thank J. Arlt for technical discussions about DDM processing, A. Brown for discussions and comments on the manuscript and acknowledge financial support from ERC-AdG-340877 PHYSAPS (V.A.M and W.C.K.P), EPSRC EP/N509620/1 (T.J.) and the Winton Programme for the Physics of Sustainability (T.J., O.A.C.).

\clearpage

%%%%%%%%%%%%%%%%%%%%%%%%%%%%%%%%%%%%%%%%%%%%%%%%%%%%%%%%%%%

\appendix

\section{Phase of the full ISF\label{sec:appendixA}}
From equation (\ref{eq:ISF}) the phase of the ISF is defined as
\be
\eta \equiv \qq\cdot\Delta\rr =\qq\cdot[\rr(\tau)-\rr(0)].
\ee
If we substitute equations (\ref{eq:rH}), (\ref{eq:drB}) and (\ref{eq:drBmag}) from the main text into (\ref{eq:qDr}) then, writing 
$\qq=q \sin\theta\,\ex + q \cos\theta\,\ez$, and noting $\ex\cdot\er=\cos\psi(\tau)$ 
and $\ex\cdot\epsi=-\sin\psi(\tau)$ (see figure \ref{fig:axes}), we obtain
\bea\label{eq:Dr00}
&&\qq\cdot\rr(\tau)= q v_p \tau \cos\theta + q R \sin\theta \cos\psi(\tau) \\\nonumber
&&+q A_b \sin(\omega_b \tau+\phi_b) \left[ \frac{v_p}{v_{h}} \cos\theta- \frac{\omega_h R}{v_{h}} \sin\psi(\tau)  \sin\theta\right].
\eea
We recall that $\psi(\tau)=\omega_h \tau + \phi_h$ is the azimuthal coordinate of the swimmer. Substituting into equation (\ref{eq:Dr00}) and setting $\tau=0$, gives $\qq\cdot\rr(0)=q R \sin\theta \cos \phi_h+q A_b \sin\phi_b \left[ \frac{v_p}{v_{h}} \cos\theta- \frac{\omega_h R}{v_{h}} \sin\phi_h  \sin\theta\right]$. Subtracting this from equation (\ref{eq:Dr00}) we arrive at expression (\ref{eq:ISFalgaederiv2}) for the ISF phase angle.

\section{Uncoupled helical and back-and-forth motions approximation\label{sec:appendixB}}

We define the small parameters $\epsilon:=\omega_h R/v_p$ and $\nu:=\omega_h/\omega_b$. Then, for the second and fourth terms in phase of the ISF (\ref{eq:ISFalgaederiv2}) in the main text, we have: $v_p/v_h =  [1+(\omega_h R/v_p)^2]^{-1/2}= 1 - \epsilon^2/2 + O(\epsilon^4)$ and, similarly, $\omega_h R/v_h = (\omega_h R/v_p) [1+(\omega_h R/v_p)^2]^{-1/2} = \epsilon - \epsilon^3/2 + O(\epsilon^5)$. Further, rescaling time such that $\tau^\prime= \omega_b \tau$, $\sin\left(\nu\tau^\prime+\phi_h\right)=\nu\tau^\prime\cos\phi_h+\sin\phi_h+O(\nu^2)$, so that equation (\ref{eq:ISFalgaederiv2}) becomes
\bea\label{eq:ISFalgaederiv2app2}
&&\eta= q v_p \frac{\tau^\prime}{\omega_b} \cos\theta + q A_b \left[\sin\left(\tau^\prime+\phi_b\right)-\sin\phi_b\right]\cos\theta\nonumber \\
&& + q R\left[\cos\left(\nu\tau^\prime+\phi_h\right)-\cos\phi_h\right] \sin\theta\nonumber\\
&& - q A_b \epsilon\sin\phi_h \left[\sin\left(\tau^\prime+\phi_b\right)-\sin\phi_b\right]\sin\theta+O(\nu\epsilon). 
\eea
If $\nu \epsilon \ll1$, the above expression for $\eta$ is identical to that in equation (\ref{eq:ISFweakcouple}) in the main text, once dimensional time is restored, and $X_b$ and $X_h$ defined in equations (\ref{eq:Xb}) and (\ref{eq:Xh}).

Next, we consider the approximation necessary to integrate (\ref{eq:ISFnobreast}) over $\phi_h$. Consider the Bessel function of the first kind $J_0(X_b)$, with argument written as $X_b=X_b^0+\delta X_b(\phi_h)$, where $X_b^0:=2 \sin\left(\frac{\omega_b \tau}{2} \right)\cos\theta$ and $\delta X_b:=-\epsilon q A_b X_b^0  \sin\phi_h \tan\theta$. Taylor expanding this function, 
\be\label{eq:XBapprox2}
J_0(X_b+\delta X_b)= J_0(X_b^0) - \delta X_b J_1(X_b^0) + O(\delta X_b^2),
\ee
where we have used $J_0^\prime=-J_1$. Thus, the integral over $\phi_h$ in (\ref{eq:ISFnobreast}) can be written as
\bea
\int_0^{2\pi}e^{i X_h \sin\left(\frac{\omega_h\tau}{2}+\phi_h\right)} J_0(X_b)d\phi_h &\approx&  J_0(X_h) J_0(X_b) \\\nonumber
&&+   \epsilon q A_b X_b^0 e^{i \pi/2} J_1(X_h)
\eea

where we have used $J_1(X_h)=(1/2\pi) \int_0^{2\pi}e^{i X_h \sin\left(\frac{\omega_h\tau}{2}+\phi_h\right)} \sin\phi_h d\phi_h$. Since $J_1(X_h)|X_b^0|\lesssim 1$, contributions from the second term to the ISF integral are negligible when $\epsilon q A_b$ is small. If $\epsilon=O(1)$ (but still $\epsilon \nu\ll1$, as above), the latter condition is still met at large lengthscales, where $q A_b\ll1$. For the parameters used in our simulations, which correspond to {\it C.reinhardtii}-like swimmers, $\epsilon\approx 2\pi f_h R/\bar{v}_p=0.2$-$0.8$ and $\nu=f_h/f_b=0.04$, $\epsilon \nu$ is always small, making equation (\ref{eq:ISFalgaederiv2app2}) a good approximation to the phase of the ISF (\ref{eq:ISFalgaederiv2}). The condition $\epsilon q A_b\ll1 $ required for the approximation to the Bessel integral just discussed, however, requires small values of the helix radius $R$ to be strictly satisfied for all $q$ values. Successful recovery of helical parameters from simulations analysed using ISF expressions for swimmers with uncoupled helical and back-and-forth motions, based on the approximations just described, further justifies the appropriateness of these approximations.
% and have defined $X_b^0=2 q A_b \sin\left(\frac{\omega_b \tau}{2} \right)\cos\theta$ and $\delta X_b= \epsilon X_b^0  \sin\phi_h \tan\theta$. Thus, if $|\delta X_b J_1(X_b^0)|\ll1$, integral (\ref{eq:helical_schulz_bstrokeandhelix}) is a good approximation to the ISF (\ref{eq:ISFnobreast}), after integrating over speeds. Recalling, the definition of $X_b^0$ above, we can write this condition as $\epsilon\, q A_b J_1(2 q A_b)\ll1$, well-satisfied if $q A_b\lesssim 1$, recalling $J_1(2)\sim1$, since $\epsilon$ is small.} 
 
 %For $q A_b\ll1$ the latter is satisfied for all values of
%$\epsilon= \omega_h R/v_p$. Otherwise, the condition is satisfied if $q A_b  \omega_h R/v_p$ is small, i.e. $q R \nu\ll1$, recalling that for swimmers	$v_p\sim \omega_b A_b$ \cite{LaugaPowers09}. This condition is satisfied for all swimmers obeying the separation of timescales assumed above.}

\section{Simulation details \label{sec:appendixC}}

Expanded in component form, equations (\ref{eq:helicalsim}) become
\begin{eqnarray} 
\dot{x}_j&=& v_j(t) \sin\theta_{p,j} \cos\phi_{p,j}\label{eq:jefftrajXS},\\
\dot{y}_j&=& v_j(t) \sin\theta_{p,j} \sin\phi_{p,j}  \label{eq:jefftrajYS},\\
\dot{z}_j&=& \cos\theta_{p,j} \label{eq:jefftrajZS},\\
\dot{\theta}_{p,j}&=&\omega_h \sin\theta_{n_0,j} \sin(\phi_{n_0,j}-\phi_{p,j}), \label{eq:jefftrajphiS}\\
\dot{\phi}_{p,j}&=&-\omega_h[\sin\theta_{n_0,j}\cos\theta_{p,j} \cos(\phi_{n_0,j}-\phi_{p,j}) - \cos\theta_{n_0} \sin\theta_{p,j}]\frac{1}{\sin\theta_{p,j}},  \label{eq:helphiS}
\end{eqnarray}
where $v_j(t)=v_{h,j} + \omega_b A_b \sin{\omega_b t}$ is the net along-helix speed including a sinusoidal contribution due to back-and-forth motions. The angles $\theta_{n_0}$ and $\phi_{n_0}$ determining the helix orientation are given by the constraint $\mathbf{p}_{j}\cdot\mathbf{n}_{0,j}=v_j/v_{h,j}=\cos \gamma_j$. This system of ODEs is solved numerically using MATLAB (Mathworks, Natick, MA, USA) with a Runge-Kutta-Fehlberg (RK45) method parallelised for $N=1000$ swimmers in a periodic box of size $L_x\times L_y \times L_z$, where $L_x=L_y=L_z=1000$ $\mu$m in figures \ref{fig:ISF_helical}, \ref{fig:ISF_bstroke}, \ref{ffig:ISF_bstrokeparams} and \ref{ffig:ISF_bstrokeandhelical}. The box size was increased to $L_x=L_y=L_z=2000$ $\mu$m in figures
\ref{fig:vq_ballistic}, \ref{fig:vq_ballisticPv} and \ref{fig:bstrokeandhelicalparams}, where the analysis of swimming parameters requires reaching lower $q$ values. In the first case, a $500$ $\mu$m thick slice of the box was taken to generate microscopy videos, whereas the full box was used in the latter case. Microscopy-like videos were generated by assigning a Gaussian pseudo-diffraction spot to the position of each simulated swimmer in the slice volume. This is achieved by subtracting the intensity $I_0\exp(-((x-x_i)^2+(y-y_i)^2)/(2\sigma^2)) [1-(z^o_i/\delta)^2]$ from the background intensity $I_B$ for of each pixel $(x,y)$ within a cut-off distance (in $x$ and $y$ only) of each alga, where $z^o_i$ is the offset of the algae from the centre of the simulation box in z-direction. Here $I_0$ is the Gaussian peak intensity, $\sigma$ its standard deviation in the plane and $\delta$ its extent in depth. The following image parameters were used in all simulations: $I_B=255$, $I_0=50$, $\delta=L_z/2$ and $\sigma=1$. The videos simulated capture at $500$ Hz, resulting in sequences of duration $15.32$ s ($L_z=1000$ $\mu$m) or $32$ s ($L_z=2000$ $\mu$m). As in the ISF model, the simulations neglect orientational noise and other realistic effects discussed in \cite{Hopeetal16}.

%\section{Supplementary figure \label{sec:appendixD}}
%
%%%%% FIGURE D1: HELICAL ISF %%%%%%%%%
%\begin{figure}[tbh!]
%   \begin{center}
%      \includegraphics[width=0.8\linewidth]{Figures/Fig_Helical_single_ISF_vp120.eps}
%      \caption{Pure helical swimming ISFs as a function of delay $\tau$ for different $q$ values from simulations of helical swimmers with a single speed (input parameters: single speed $v^1_h=120$ $\mu$/s, helical frequency $f_h=2$ Hz and radius $R_h=8$ $\mu$m), noting for these results it is the along-helix speed that is fixed. There is no obvious signature of the helical path (\textit{no kink}), presumably because the ballistic time-scale is faster and the \textit{kink} is in the noise floor. %\blue{\\TO DO:  
%   %\begin{itemize}
%     %\item a, b and d: add labels
%     %\item b, d: add units (display as you do for q values in a))
%     %\item a: remove parameters from legend
%     %\item d: just quote $\sigma_p$ (adjust legend values)
%%      \item plot for (d) might be replaced simply by a plot of the ISFs at one q=0.063 and R=8, but changing Z:5,10,20, single speed: this should show the disappearance of the "kink". NOT SURE WE NEED TO DO THIS.
%     %\end{itemize}}
%      }
%      \label{fig:ISF_helical_v1h}
%   \end{center}
%\end{figure}
%

\section*{References}

\bibliographystyle{unsrt}

\providecommand{\noopsort}[1]{}\providecommand{\singleletter}[1]{#1}%

\end{document}